# Thermal conductivities of phosphorene allotropes from first-principle calculations: a comparative study

J. Zhang, H. J. Liu[*], L. Cheng, J. Wei, J. H. Liang, D. D. Fan, P. H. Jiang, J. Shi

*Key Laboratory of Artificial Micro- and Nano-Structures of Ministry of Education and School of Physics and Technology, Wuhan University, Wuhan 430072, China*

Phosphorene has attracted tremendous interest recently due to its intriguing electronic properties. However, the thermal transport properties of phosphorene, especially for its allotropes, are still not well-understood. In this work, we calculate the thermal conductivities of five phosphorene allotropes (α-, β-, γ-, δ- and ζ-phase) by using phonon Boltzmann transport theory combined with first-principles calculations. It is found that the α-phosphorene exhibits considerable anisotropic thermal transport, while it is less obvious in the other four phosphorene allotropes. The highest thermal conductivity is found in the β-phosphorene, followed by δ-, γ- and ζ-phase. The much lower thermal conductivity of the ζ-phase can be attributed to its relatively complex atomic configuration. It is expected that the rich thermal transport properties of phosphorene allotropes can have potential applications in the thermoelectrics and thermal management.

**Introduction**

Single-layer black phosphorus, the so-called phosphorene, has emerged as a viable candidate in the field of two-dimensional atomic-layer materials due to its high carrier mobility and a thickness-dependent direct band gap. Besides phosphorene, several other stable layered phosphorene allotropes, with similar buckled or puckered honeycomb structure have been theoretically predicted by Zhu *et al.* [1, 2]. It is suggested that phosphorene with distinct configurations possess different electronic properties, which can be further modulated by the layer thickness, in-layer strain, and heterostructure assembling without energy penalty. Such tunable electronic properties of phosphorene allotropes are very beneficial for optoelectronic and nanoelectronic applications. Although high performance field-effect transistors (FET) and photovoltaic devices based on phosphorene have been reported recently [3, 4, 5, 6, 7], the anisotropic thermal transport properties of phosphorene may degrade the device reliability and performance since the low thermal conductivity along the armchair direction can lead the localized Joule heating in the confined system. On the other hand, the possibility to use phosphorene as thermoelectric material has been theoretically suggested [8, 9]. For example, Lv *et al.* showed that the power factor of phosphorene can reach as high as 138.9 $\mu Wcm^{-1}K^{-2}$ at appropriate carrier concentration. However, they predicted that the thermoelectric performance of phosphorene is poor at room temperature due to relatively high lattice thermal conductivity, implying that suppressing thermal transport is an efficient way to enhance its thermoelectric efficiency. Although all-scale hierarchical architecturing and nanostructuring are efficient approaches to reduce the thermal conductivity of thermoelectric materials, the corresponding synthetic techniques are usually complicated and costly. As an alternative, high thermoelectric performance could be sought in materials with intrinsically low thermal conductivity [10]. In order to apply phosphorene allotropes as promising nanoelectronics and thermoelectric materials, it is necessary to investigate their thermal transport properties.

---
[*] Author to whom correspondence should be addressed. Electronic mail: phlhj@whu.edu.cn.

In this work, using phonon Boltzmann transport theory combined with first-principles calculations, we provide a comparative study on the thermal transport properties of five phosphorene allotropes. We demonstrate that thermal transport in the α-phosphorene exhibits strong orientation dependence, which is less obvious in the β-, γ-, δ- and ζ-phase. Moreover, we find that the ζ-phosphorene possesses much lower thermal conductivity, which is consistent with its relatively complex atomic configuration. Our theoretical work not only offers physical insight into the thermal transport in phosphorene allotropes, but also suggests that low thermal conductivity can be achieved in crystalline systems constructed by light phosphorus atoms.

**Methods**

Our theoretical calculations are performed by using a first-principles plane-wave pseudopotential formulation [11, 12, 13] as implemented in the Vienna *ab-initio* (VASP) [14] code. The exchange-correlation functional is in the form of Perdew-Burke-Ernzerhof (PBE) [15] with the generalized gradient approximation (GGA). For the structural optimization, the energy convergence threshold is set to $1\times10^{-7}$ eV and the residual force on each atom is less than $10^{-5}$ eV/Å. The cutoff energy for the plane wave basis is set to be 500 eV, and uniform Monkhorst-Pack [16] *k*-meshes are applied to sample the Brillouin zone. To eliminate interactions between the phosphorene layer and its periodic images, we use a vacuum distance larger than 14 Å for the supercell geometry.

The lattice thermal conductivity of phosphorene allotropes can be calculated by using phonon Boltzmann transport equation (BTE) with relaxation time approximation, as implemented in the so-called ShengBTE code [17, 18, 19]. In this approach, the thermal conductivity along the $\alpha$ direction can be calculated by:

$$\kappa_\alpha = \frac{1}{N_q V} \sum_{\vec{q},j} C_{\vec{q},j} v^2_{\vec{q},j,\alpha} \tau_{\vec{q},j}. \qquad (1)$$

Here $C_{\vec{q},j}$ is the specific heat contribution of the phonon mode with the wave vector $\vec{q}$ and polarization $j$, $v_{\vec{q},j,\alpha}$ is the corresponding phonon group velocity, and $\tau_{\vec{q},j}$ is the phonon relaxation time. $N_q$ is the number of sampled *q* points in the Brillouin zone, and *V* is the volume of the unit cell. During the calculations of thermal conductivity, the only inputs are the harmonic and anharmonic interatomic force constants (IFCs) matrix, which can be extracted from first-principles calculations. To calculate the second- and third-order IFCs, a 6×6, 5×5, 5×4, 3×3, and 3×3 supercell is respectively used for the α-, β-, γ-, δ- and ζ-phosphorene. When dealing with the anharmonic ones, a cutoff distance of about 5.5 Å is employed. It should be mentioned that the phonon BTE method has already been used to predict the lattice thermal conductivity of many bulk and low-dimensional structures such as transition-metal dichalcogenides, and the results are in good agreement with the experimental measurements [20, 21, 22].

**Results and discussion**

The crystal structures of phosphorene allotropes are schematically depicted in Figure 1. Following the well-known notation for phosphorene allotropes [2], we denote them as α-phosphorene (Fig. 1(a)), β-phosphorene (Fig. 1(b)), γ-phosphorene (Fig. 1(c)), δ-phosphorene (Fig. 1(d)), and ζ-phosphorene (Fig. 1(e)), respectively. One can see that all the phases share the structural motif of threefold-coordinated P atoms and characterized by a non-planar honeycomb atomic-layer. There are four atoms in the rectangular unit cell of α- and γ-phosphorene, and eight atoms in that of δ- and ζ-phosphorene. For the isotropic β-phase, however, there are only two P atoms in the hexagonal unit cell. The optimum structural parameters for the five phosphorene allotropes are summarized in Table I. It should be noted that our calculated results for the α-, β-, γ- and δ-phase are very close to previous first-principles calculations [2, 23]. For the less studied ζ-phase, we see it has larger lattice constant and more atoms in the unit cell compared with those of other phases, suggesting that P atoms in the ζ-phosphorene are arranged in a more complicated way. Such behavior would set a crucial precondition for the ζ-phase to exhibit a lower thermal conductivity, as will be discussed in the following. To investigate the stability of the phosphorene phases, we have computed the energy difference ($\Delta E$) with respect to the most stable α-phosphorene by using the formula:

$$\Delta E = E - E(\alpha - \text{phase}), \qquad (2)$$

where $E$ is the total energy per atom of the phosphorene allotropes. As can be found from Table I, the β-phase is almost as stable as the α-phase, while the γ- and δ-phosphorene have higher energies than that of the α-phase by 95 and 91 meV/atom, respectively. Although the energy difference between the ζ- and α-phosphorene is 135 meV/atom, we can still expect that the ξ-phosphorene is a stable phase by considering the fact that $\Delta E$ is lower than the energy difference between black and white phosphorus (160 meV/atom) [24]. To further confirm the thermal stability of the ζ-phosphorene, we have performed *ab-initio* molecular dynamics (MD). The system is simulated in a microcanonical ensemble for 1000 steps with a time step of 1.0 fs. Figure 2 shows the structural snap shots of ζ-phosphorene at 300, 500, and 700 K. We see that the structure of ζ-phase has slight fluctuations even at 700 K, which indicates that the ζ-phosphorene considered in our work is rather stable.

Figure 3 plots the phonon dispersion relations of the five phosphorene allotropes. For the well-known α-, β-, γ- and δ-phase, our results are found to be consistent with recent works by using the finite displacement method [1, 2, 25] and further confirms the reliability of our calculations. For the less-studied ζ-phosphorene, we see there is no imaginary frequency in the phonon spectrum which suggests that it is also kinetically stable. As indicated in the figure, the α- and β-phosphorene have obviously larger phonon gaps than the other three phases, which suggests that the rate for three-phonon scattering is lower by considering the energy conservation. It is thus reasonable to expect that the α- and β-phosphorene may have a relatively higher thermal conductivity, as will be discussed later. Moreover, we see that the phonon dispersion of α-phosphorene is highly asymmetric along the Γ-X and Γ-Y directions, which is caused by its puckered hinge-like structure [26, 27]. In contrast, such behavior is less obvious for the other four phosphorene allotropes. In Table II, we list the Γ point group velocities of longitudinal acoustic (LA) phonons for the five phosphorene allotropes. We find that the group velocity of α-phosphorene along the *x*-direction is two times higher than that along the *y*-direction, which suggests that the thermal transport of α-phosphorene will have significant orientation dependence. For the other four phosphorene allotropes, however, the differences between *x*- and *y*-direction are relatively small (it is identical for the β-phosphorene), implying their lattice thermal conductivity will exhibit less anisotropy.

Figure 4 shows the calculated lattice thermal conductivity ($\kappa_p$) of five phosphorene allotropes as a function of temperature in the range from 200 K to 800 K. As the value of $\kappa_p$ for two-dimensional system is related to the layer thickness, we adopt the interlayer separation of the bulk counterparts from Reference [2], which is 5.30, 4.20, 4.21 and 5.47 Å for the α-, β-, γ- and δ-phase, respectively. For the less-studied ζ-phase, we apply the same method as used in the Reference [2] to optimize the corresponding bulk structure, and get a interlayer separation of 4.89 Å. It can be seen that for all the five phosphorene allotropes, the $\kappa_p$ decrease with increasing temperature and roughly follow a $T^{-1}$ dependence, indicating that the Umklapp process is the dominant phonon scattering mechanism in the temperature range considered. In general, the thermal conductivities of four almost isotropic phosphorene allotropes are in the order of β-phase > δ-phase > γ-phase > ζ-phase. For the α-phosphorene, we see it indeed exhibits obvious anisotropic thermal transport. For example, the room temperature thermal conductivity along the *x*-direction is almost four times higher than that along the *y*-direction. Such behavior is consistent with a recent work carried out by using first-principles calculations [25, 26] and molecular dynamics simulations [28]. Among the five investigated phosphorene allotropes, it is very hopeful to use β-phase to solve the thermal management issues in the phosphorene based FETs. In addition, a similar lattice structure and thermal conductivity to MoS$_2$ monolayer will make β-phosphorene a promising material to form phosphorene/MoS$_2$ superlattice, whose electronic properties is highly tunable by controlling the layer thickness and stacking order. On the other hand, the thermal conductivity of ζ-phosphorene can be as small as 2.7 W/mK at 800 K, which is rare for crystalline system constructed by light atoms. Combined with good electronic transport properties [ 29 ], it is very promising to use ζ-phosphorene as high-performance thermoelectric material.

It should be mentioned that the calculated thermal conductivity of α-phosphorene is much lower than that of graphene (2000 ~ 5000 W/mK). The reason is that the no-planar structure of α-phosphorene breaks the out-of-plane symmetry and promotes the phonon-phonon scattering of the out-of-plane acoustic (ZA) mode [26, 27]. As can be seen from Table III, the contribution of ZA phonons to the thermal conductivity in the α-phosphorene is much smaller than that of graphene (76%) [30]. In addition, we see the contribution of ZA mode is smaller than that of transverse acoustic (TA) and longitudinal acoustic (LA) mode in the β- and γ-phosphorene. Moreover, except for the β-phase, all the other four allotropes have a relatively higher optical contribution, which may arise from the hybridization of low frequency optical phonons with the longitudinal acoustic phonons (see Fig. 3). Such behavior will flatten the LA dispersion and induce suppression of thermal transport from LA phonons [31]. In this regard, neglecting the optical-phonon contributions will result in an underestimated thermal conductivity [32]. Surprisingly, the contribution from the optical mode is higher than 30% in the ζ-phosphorene, which can be attributed to its more complex atomic configuration. Similar behavior has also been found in the complex compounds such as SnSe and $CoSb_3$ skutterudite [33, 34].

To have a better understanding of the calculated thermal conductivity of the five phosphorene allotropes, we plot in Figure 5 the accumulative thermal conductivity at room temperature with respect to cutoff phonon mean free path (MFP). For all the phosphorene allotropes, one can see that heat is mainly carried by phonons with MFP in a broad range (in the orders of magnitude from 10 to $10^3$ nm). To figure out which kind of phonons can have a significant effect, we give in Table IV the MFP values corresponding to 50% $\kappa_p$ accumulation. It can be seen that such characteristic MFP for the β- and δ-phosphorene have an order of magnitude of about 100 nm, which means that the lattice thermal conductivity could be effectively decreased if the sample size is smaller than this critical value. For the ζ-phosphorene, however, the major contribution comes from phonons with MFP of about 10 ~ 20 nm, which is similar to that found in silicene where phonons with MFP of 5 ~ 20 nm contribute more than 80% of the thermal conductivity [35]. It is interesting to note that the characteristic MFP of the γ-phase exhibits strong anisotropy, which offers extra flexibility to modulate its thermal conductivity along the *y*-direction.

To understand why the five phosphorene allotropes have quite different thermal conductivity, we first examine their group velocities of different phonon branches. As shown in Figure 6, we see that the group velocities of optical phonons is comparable with that of acoustic phonons in the α-, γ-, δ- and ζ-phosphorene, which is consistent with the fact that optical phonons contribute considerably to the total thermal conductivity of these four phases. For the acoustic phonons, we find that the group velocity of LA and ZA modes in the α-phase and LA and TA modes in the β-phase are relatively higher than those in the other three phases, which may result in a higher thermal conductivity of these two phases. We next focus on the phonon relaxation time ($\tau$) of these phosphorene allotropes. As can be seen from Figure 7, the β-phase has smaller relaxation time of optical phonon compared with those of other phases. Moreover, the lowest relaxation time for acoustic phonons can be found in the ζ-phosphorene, especially for the TA and LA modes. As a result, the ζ-phase exhibits the lowest thermal conductivity among the five investigated phosphorene allotropes. To further estimate the phonon-phonon scattering, we plot in Figure 8 the dimensionless scattering phase space (the so-called P3 parameter). It is clear to find that the γ- and ζ-phosphorene possess larger scattering phase space in the whole frequency range considered, especially for the case of acoustic phonons. This means that there are more phase space allowing the phonon-phonon scattering in the γ- and ζ-phosphorene, which can lead to a reduction of phonon relaxation time and thus the lattice thermal conductivity.

**Conclusion**

We demonstrate by phonon Boltzmann theory combined with first-principles calculations that thermal transport in the α-phosphorene exhibits strong orientation dependence, which is less obvious for β-, γ-, δ- and ζ-phase. Moreover, we find that the less-studied ζ-phosphorene exhibits considerably small thermal conductivity in a broad temperature range, which is very desirable for thermoelectric application. Detailed analysis of the characteristic MFP of these phosphorene allotropes suggests that the thermal conductivity of β- and δ-phosphorene could be modulated effectively by controlling the sample size. In contrast to the general assumption, we find that the optical modes contribute significantly to the total thermal conductivity (except for the β-phosphorene) and thus cannot be ignored. On the experimental side, considering the fact that α-phosphorene has been synthesized using mechanical exfoliation method [3, 6, 36], it is possible that other four phosphorene allotropes can be obtained in a similar way [1, 2]. In addition, molecular beam epitaxy (MBE) and chemical vapor deposition (CVD) may be alternative approaches to produce these two-dimensional materials [1, 37]. With the rapid progress of fabrication techniques, it is reasonable to expect that thermoelectric devices and thermal management in nanoelectronics could be realized in the phosphorene allotropes.

## Acknowledgements

We thank financial support from the National Natural Science Foundation (Grant No. 11574236 and 51172167) and the "973 Program" of China (Grant No. 2013CB632502).

**Table I** The optimized lattice constants ($a_1$, $a_2$), total energy ($E$), and energy relative to the α-phase ($\Delta E$) of the five phosphorene allotropes.

| Phase | α | β | γ | δ | ξ |
|---|---|---|---|---|---|
| $a_1$ (Å) | 3.30 | 3.28 | 3.27 | 5.54 | 5.78 |
| $a_2$ (Å) | 4.62 | 3.28 | 5.53 | 5.64 | 6.22 |
| $E$ (eV/atom) | −5.364 | −5.362 | −5.269 | −5.273 | −5.229 |
| $\Delta E$ (eV/atom) | 0 | 0.002 | 0.095 | 0.091 | 0.135 |

**Table II** The Γ point group velocity of longitudinal acoustic phonons along the *x*- and *y*-direction for the five phosphorene allotropes.

| Velocity (km/s) | α | β | γ | δ | ζ |
|---|---|---|---|---|---|
| *x*-direction | 8.58 | 8.05 | 8.14 | 5.40 | 7.83 |
| *y*-direction | 4.18 | 8.05 | 7.02 | 6.19 | 5.72 |

**Table III** Percentage contribution of different phonon branches to the room temperature thermal conductivity for the five phosphorene allotropes.

| Contribution (%) | α | β | γ | δ | ζ |
|---|---|---|---|---|---|
| ZA-*x* | 29.1 | 19.1 | 10.9 | 9.0 | 42.6 |
| ZA-*y* | 7.7 | | 5.2 | 31.6 | 18.0 |
| TA-*x* | 19.9 | 43.7 | 35.6 | 33.5 | 15.7 |
| TA-*y* | 14.0 | | 74.6 | 27.3 | 18.4 |
| LA-*x* | 35.9 | 34.3 | 20.4 | 35.5 | 10.4 |
| LA-*y* | 34.7 | | 9.5 | 27.6 | 15.1 |
| Optical-*x* | 15.1 | 2.9 | 33.1 | 22.0 | 31.3 |
| Optical-*y* | 43.6 | | 10.7 | 13.5 | 48.5 |

**Table IV** The MFP values corresponding to 50% $\kappa_P$ accumulation for the five phosphorene allotropes.

| MFP (nm) | α | β | γ | δ | ζ |
| --- | --- | --- | --- | --- | --- |
| $x$-direction | 103 | 146 | 40 | 446 | 23 |
| $y$-direction | 37 | 146 | 2138 | 583 | 11 |

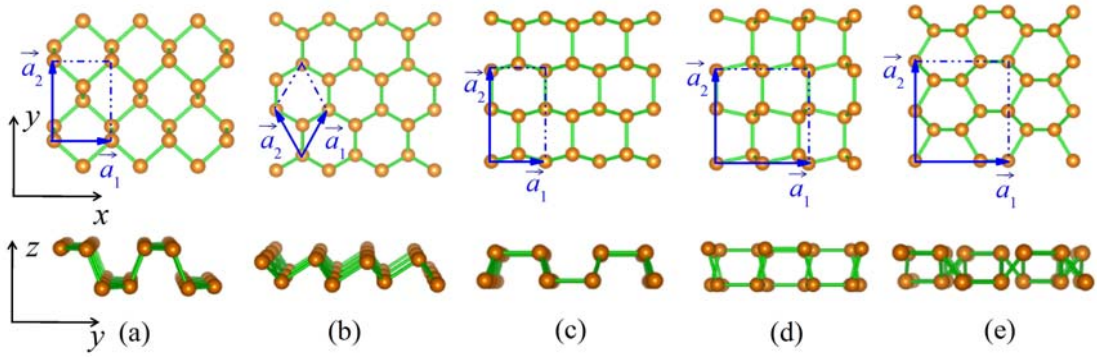

**Figure 1** The top- and side-views of (a) α-phosphorene, (b) β-phosphorene, (c) γ-phosphorene, (d) δ-phosphorene, and (e) ζ-phosphorene. The coordinate axes ($x$, $y$, $z$) and lattice vectors ($\vec{a_1}$, $\vec{a_2}$) are indicated.

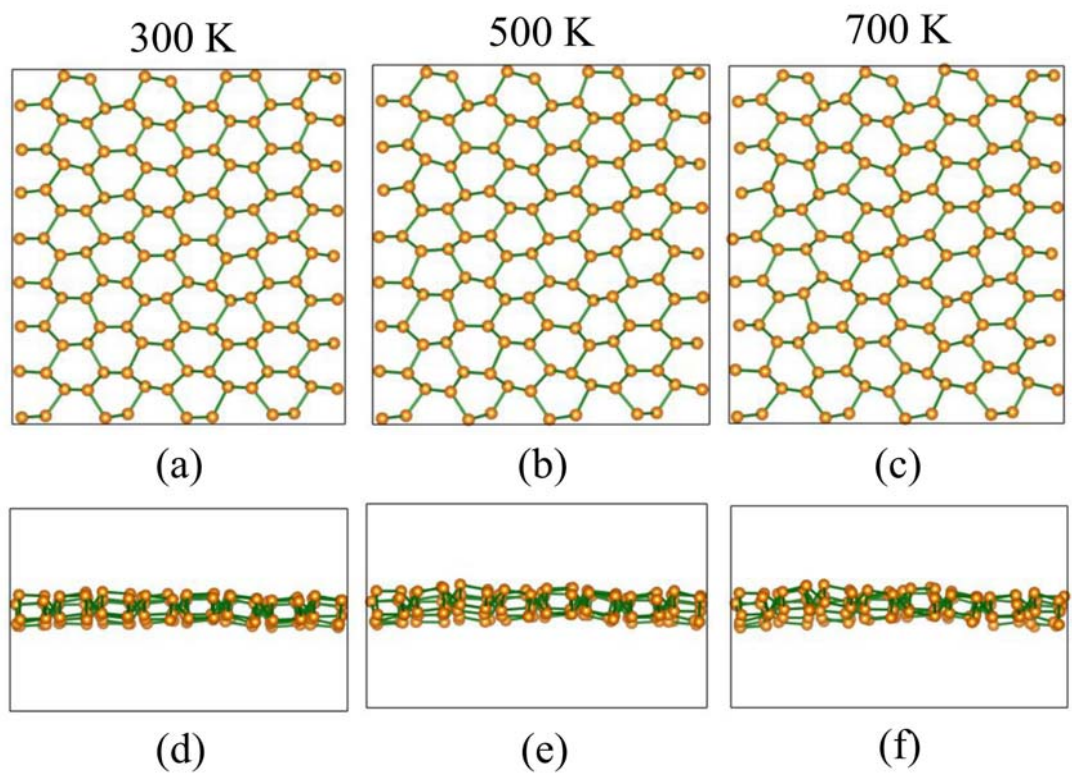

**Figure 2** Structural snap shots of the ζ-phosphorene at (a) 300 K, (b) 500 K, and (c) 700 K during the molecular dynamics simulations. The corresponding side-views are shown in (d), (e), and (f), respectively.

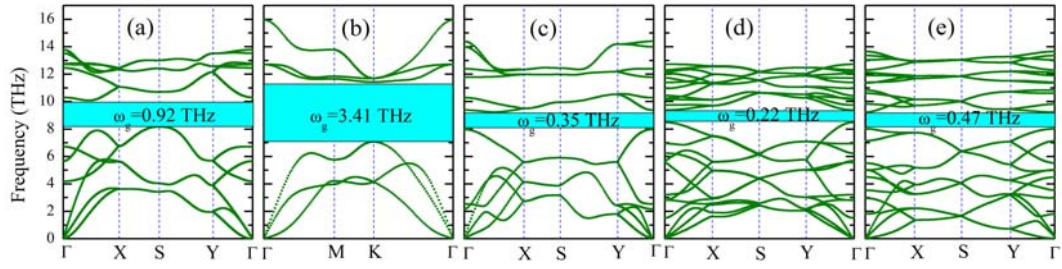

**Figure 3** The phonon dispersion relations of (a) α-phosphorene, (b) β-phosphorene, (c) γ-phosphorene, (d) δ-phosphorene, and (e) ζ-phosphorene. The corresponding phonon gap is highlighted.

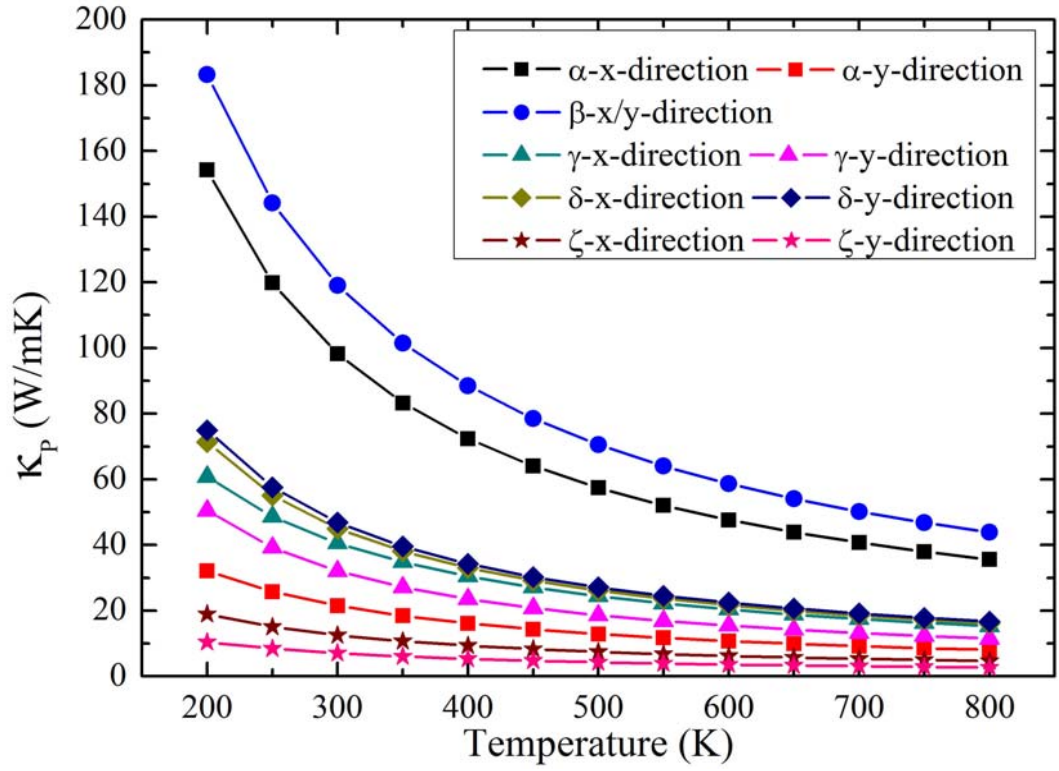

**Figure 4** Calculated thermal conductivity of five phosphorene allotropes as a function of temperature.

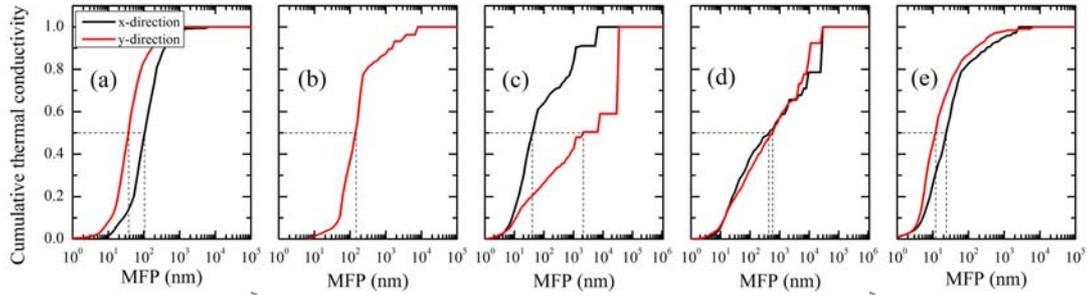

**Figure 5** The accumulative thermal conductivity at room temperature as a function of cutoff phonon MFP for: (a) α-phosphorene, (b) β-phosphorene, (c) γ-phosphorene, (d) δ-phosphorene, and (e) ζ-phosphorene. The dashed line denotes the MFP values corresponding to 50% $\kappa_P$ accumulation.

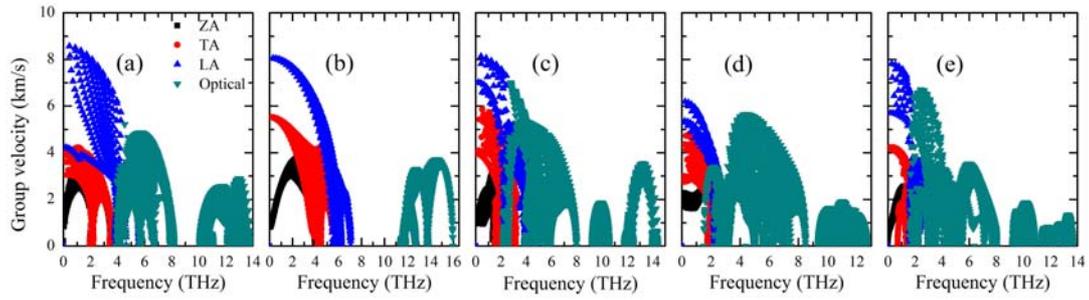

**Figure 6** Room temperature group velocities of different phonon modes as a function of frequency for: (a) α-phosphorene, (b) β-phosphorene, (c) γ-phosphorene, (d) δ-phosphorene, and (e) ζ-phosphorene.

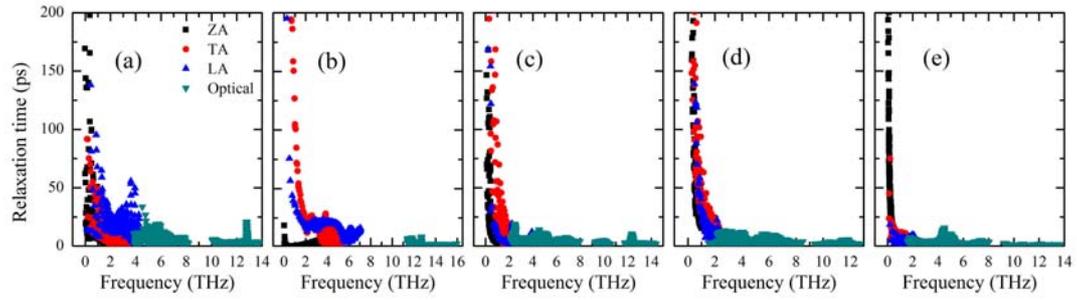

**Figure 7** Room temperature relaxation time of different phonon modes as a function of frequency for: (a) α-phosphorene, (b) β-phosphorene, (c) γ-phosphorene, (d) δ-phosphorene, and (e) ζ-phosphorene.

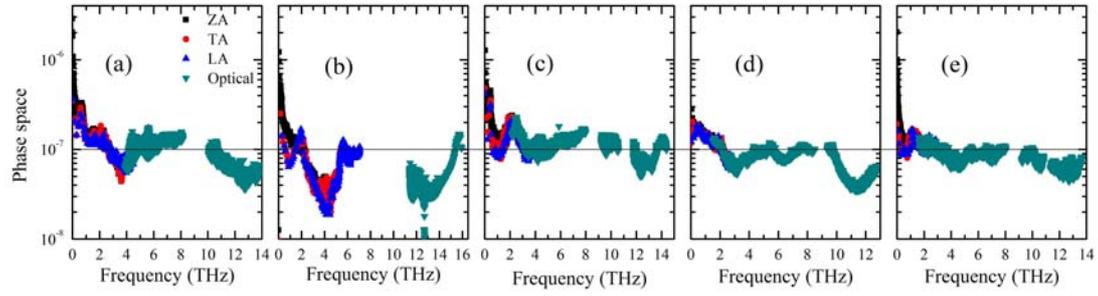

**Figure 8** Phase space of different phonon modes in three-phonon scattering processes for: (a) α-phosphorene, (b) β-phosphorene, (c) γ-phosphorene, (d) δ-phosphorene, and (e) ζ-phosphorene. The black line corresponds to a phase space of $10^{-7}$ and is shown to guide the eyes.